# Coverage and adoption of altmetrics sources in the bibliometric community

Stefanie Haustein[1], Isabella Peters[2], Judit Bar-Ilan[3], Jason Priem[4], Hadas Shema[3], and Jens Terliesner[2]

[1] *stefanie.haustein@umontreal.ca*
École de bibliothéconomie et des sciences de l'information, Université de Montréal, Montréal (Canada)
and Science-Metrix, 1335 Avenue Mont-Royal Est, Montréal, H2J 1Y6 (Canada)

[2] *isabella.peters@hhu.de; jens.terliesner@hhu.de*
Department of Information Science, Heinrich-Heine-University, Universitätsstr. 1, Düsseldorf, 40225 (Germany)

[3] *Judit.Bar-Ilan@.biu.ac.il; dassysh@gmail.com*
Department of Information Science, Bar-Ilan University, Ramat-Gan, 52900 (Israel)

[4] *priem@email.unc.edu*
School of Information & Library Science, University of North Carolina at Chapel Hill, 216 Lenoir Drive, CB #3360100 Manning Hall, Chapel Hill (USA)

**Abstract**
Altmetrics, indices based on social media platforms and tools, have recently emerged as alternative means of measuring scholarly impact. Such indices assume that scholars in fact populate online social environments, and interact with scholarly products there. We tested this assumption by examining the use and coverage of social media environments amongst a sample of bibliometricians. As expected, coverage varied: 82% of articles published by sampled bibliometricians were included in Mendeley libraries, while only 28% were included in CiteULike. Mendeley bookmarking was moderately correlated (.45) with Scopus citation. Over half of respondents asserted that social media tools were affecting their professional lives, although uptake of online tools varied widely. 68% of those surveyed had LinkedIn accounts, while Academia.edu, Mendeley, and ResearchGate each claimed a fifth of respondents. Nearly half of those responding had Twitter accounts, which they used both personally and professionally. Surveyed bibliometricians had mixed opinions on altmetrics' potential; 72% valued download counts, while a third saw potential in tracking articles' influence in blogs, Wikipedia, reference managers, and social media. Altogether, these findings suggest that some online tools are seeing substantial use by bibliometricians, and that they present a potentially valuable source of impact data.

**Conference Topic**
Old and New Data Sources for Scientometric Studies: Coverage, Accuracy and Reliability (Topic 2); Webometrics (Topic 7)

**Introduction**
Altmetrics, indices based on activity in social media environments, have recently emerged as alternative means of measuring scholarly impact (Priem, 2010; Priem et al., 2010). The idea of impact measuring which moves beyond citation analysis, however, emerged long before the advent of social media (Martin & Irvine, 1983; Cronin & Overfelt, 1994). One of the underlying problems with citation analysis as basis for evaluating scientific impact is that citations paint a limited picture of impact (Haustein, in press). On the one hand, researchers often fail to cite all influences (MacRoberts & MacRoberts, 1989). On the other hand, the total readership population includes not only authors but also "pure," i.e. non-publishing, readers, who are estimated to constitute one third of the scientific community (Price & Gürsey, 1976; Tenopir & King, 2000). Publications are used in the development of new technologies, applied in daily work of professionals, support teaching, and have other societal effects (Schlögl & Stock, 2004; Rowlands & Nicholas, 2007; Research Councils UK, 2011; Thelwall, 2012).

Thus, a better way of approaching scholarly impact is to consider citation as just one in a broader spectrum of possible uses. Webometrics and electronic readership studies gathered impact and usage data in a broader sense, but have been restricted by scalability problems and access to data (Thelwall, Vaughan, & Björneborn, 2005; Thelwall, 2010). As altmetrics are based on clearly defined social media platforms, that often provide free access to usage data through Web APIs, data collection is less problematic, although accuracy is still a problem (Priem, in press). With these new sources comes the possibility of analyzing online usage of scholarly resources independently of publishers. Tracking the use of scholarly content in social media means that researchers are able to analyze impact more broadly (Li, Thelwall, & Guistini, 2012; Piwowar, 2013). Moreover, many online tools and environments surface evidence of impact relatively early in the research cycle, exposing essential but traditionally invisible precursors like reading, bookmarking, saving, annotating, discussing, and recommending articles.

In order to explore the potential of altmetrics, this work studies the applicability and use of altmetrics sources and indicators in the bibliometric community. As it is still unclear how broadly these platforms are used, by whom and for what purposes, this study aims to evaluate the representativeness and validity of altmetrics indicators using the bibliometric community and literature as an initial reference set. We focus on measuring the impact of conventional peer-reviewed publications, such as journal articles and proceedings papers, on the social web as well as how bibliometricians perceive and use social media tools in their daily work routine. New forms of output, such as research results published in blogs, comments and tweets, are not addressed in this paper.

We apply a two-sided approach, aiming to answer the following sets of research questions:
- RQ 1: To what extent are bibliometrics papers present on social media platforms? How comprehensive is the coverage of the literature on platforms like Mendeley and CiteULike? How many users do they have and how many times are they used?
- RQ 2: To what extent is the bibliometric community present on social media platforms? Who uses these platforms and for what purposes?

We answered the first set of questions by evaluating the coverage and intensity of use of bibliometrics literature in social reference managers. Publications by presenters of the 2010 STI conference served as a reference set, as they represent a group of both established and new bibliometricians. The second set of research questions was approached by surveying the attendees of the 2012 STI conference in Montréal regarding their use of social media.

**Altmetrics Literature Review**

Altmetrics research to date has focused on exploring potential data sources, correlating alternative impact data with citations and analyzing it from a content perspective; for overviews of this research see Bar-Ilan, Shema, and Thelwall (in press), Haustein (in press), and Priem (in press). When it comes to monitoring the impact of scholarly publications, Mendeley (mendeley.com) and CiteULike (citeulike.org) have proven particularly useful. They combine social bookmarking and reference management functionalities and allow users to save literature, share them with other users, and add keywords and comments (Henning & Reichelt, 2008; Reher & Haustein, 2010). Both social bookmarking systems use a bag model for resources, meaning that a particular resource can be simultaneously saved or bookmarked by several users. This functionality allows for counting resource-specific bookmarking actions like how many users saved a particular resource. According to self-reported numbers, Mendeley is considerably larger than CiteULike (CuL). During data collection in March 2012, CuL claimed to have 5.9 million unique papers in CuL vs. more than 34 million in

Mendeley (Bar-Ilan, Haustein, Peters, Priem, Shema, & Terliesner, 2012). As of August 2012, Mendeley claims to be the largest research catalog with 280 million bookmarks to 68 million unique documents uploaded by 1.8 million users (Ganegan, 2012). In November 2012 Mendeley reached 2 million users (Mendeley, 2012).

Case studies focusing on the coverage of social reference managers support Mendeley's position as a leader in the field. Li, Thelwall, and Giuistini (2012) investigated how bookmarks in Mendeley and CuL reflect papers' scholarly impact and found that 92% of sampled Nature and Science articles had been bookmarked by at least one Mendeley user, and 60% by one or more CuL users. Bar-Ilan (2012a; 2012b) found 97% coverage of recent JASIST articles in Mendeley. Priem, Piwowar, and Hemminger (2012) showed that the coverage of articles published in the PLoS journals was 80% in Mendeley and 31% in CuL. Li and Thelwall (2012) sampled 1,397 F1000 Genomics and Genetics papers and found that 1,389 of those had Mendeley users.

Studies have found moderate correlation between bookmarks and Web of Science (WoS) citations. Li, Thelwall, and Giustini (2012) reported r=.55 of Mendeley and r=.34 of CuL readers with WoS citations, respectively. Weller and Peters (2012) arrived at slightly higher correlation values for a different article set between Mendeley, CuL, BibSonomy, and Scopus. Bar-Ilan (2012a; 2012b) found a correlation of .46 between Mendeley readership counts and WoS citations for the JASIST articles. Li and Thelwall (2012) found high correlation (.69) between Mendeley and WoS for the articles recommended on F1000. User-citation correlations for the Nature and Science publications were .56 (Li, Thelwall, & Guistini, 2012) and Priem, Piwowar, and Hemminger (2012) found a correlation of .5 between WoS citations and Mendeley users for the PLoS publications.

While bookmarks in reference managers reflect readership of scholarly articles, Twitter activity reflects discussion around these articles. Several studies have analyzed tweets "citing" scholarly publications. Priem and Costello (2010) and Priem, Costello, and Dzuba (2011) found that scholars use Twitter as a professional medium for sharing and discussing articles, while Eysenbach (2011) showed that highly-tweeted articles were 11 times more likely become highly-cited later. Weller and Puschmann (2011), and Letierce, Passant, Decker, and Breslin (2010) analyzed the use of Twitter during scientific conferences and revealed that there was discipline-specific tweeting behavior regarding topic and number of tweets as well as references to different document types (i.e., blogs, journal articles, presentation slides). Along with Twitter, other studies have examined citation from Wikipedia articles (Nielsen, 2007) and blogs (Groth & Gurney, 2010; Shema, Bar-Ilan, & Thelwall, 2012) as potential sources reflecting alternative impact of scholarly documents.

Apart from aforementioned studies, which focused on quantitative analysis of social media impact, there is a more content-oriented research approach which particularly examines tags attached to products of scholarly practice. Bar-Ilan (2011) studied the items tagged with "bibliometrics" on Mendeley and CuL, whereas Haustein and Peters (2012) and Haustein et al. (2010) showed that tags represent a reader-specific view on articles' content which could be used to analyze journal content from a readers perspective (as opposed to the author and indexer perspectives).

Although altmetric indicators and data sources are increasingly applied in evaluation studies, little is yet known about the users of such social media platforms or how researchers integrate them into their research environment (Mahrt, Weller, & Peters, in press). Understanding who is using social media tools for which purpose is, however, crucial to the application of altmetrics for evaluation purposes. Given that a representative share of documents are covered by social media tools and the user community can be identified, social media platforms can be valuable sources for measuring research impact from the readers' point of view, functioning as supplements to citation analysis. In contrast to citations, altmetrics potentially cover the whole readership and are available in real time.

**RQ 1: Coverage of Bibliometrics Papers on Altmetrics Platforms**

Before analyzing the alternative impact of bibliometrics literature and authors from the bibliometric community, it is necessary to explore which sources are suitable and provide the best coverage. Comparing them to traditional sources of impact evaluation provides information about the differences between use in citation and use in other contexts.

*Method*

In order to create a list of bibliometrics publications, all documents authored by presenters of the 2010 STI conference in Leiden were collected on WoS and Scopus. We chose this author-based, bottom-up approach to facilitate linking altmetrics data to authors as well as just documents. The group of presenters at the STI conference was considered to represent a core group of both established and new members of the current bibliometric community. The presenters' names were retrieved from the conference program. The final list contained 57 researchers, who together had authored 1,136 papers[1] covered in Scopus. Mendeley publication and readership information was retrieved manually via the Mendeley Web search interface from mendeley.com. At the time of data collection in March 2012 the manual approach proved more comprehensive, as the API, searched via the ImpactStory tool[2], only returned one of multiple entries matching the search criteria. More recent searches seem to indicate this problem has since been resolved. In CuL, publications can be searched by DOI. However, it should be noted that bibliographic data in CuL or Mendeley is incomplete (Haustein & Siebenlist, 2011). The number of articles bookmarked in CuL might thus be higher than the number retrieved via DOI. The manual search in Mendeley showed that 33% of the documents retrieved did not contain a DOI.

*Results*

As shown in Table 1, the coverage of the 1,136 bibliometrics documents in Mendeley was good: 928 (82%) of the documents had at least one Mendeley bookmark, while only 319 (28%) of articles were in CuL. Although coverage in CuL may be underestimated because bookmarks without a correct DOI were not retrieved, this confirms the results found by other studies (e.g., Li, Thelwall, & Guistini, 2012; Priem, Piwowar, & Hemminger, 2012). Unsurprisingly given Mendeley's very recent founding, older articles are less bookmarked. Of the 85 sample articles published before 1990, only 44% have readers in Mendeley, while 88% of those published since 2000 have Mendeley bookmarks (see Figure 1). Mendeley's popularity is not only reflected in the coverage of documents but also by the average activity on bookmarked documents: in Mendeley each document was bookmarked by a mean of 9.5 users, compared to a usage rate of 2.4 in CuL. Correlations between Scopus citations and users counts were .45 for Mendeley and .23 for CuL. These moderate correlations confirm previous findings for other samples and suggest that altmetrics may indeed reflect impact not reflected in citation counts.

---

[1] Some presenters were omitted either because they had not published in sources covered by Scopus or WoS or due to ambiguous names, for which relevant papers could not be identified. Documents without a DOI were not considered as it was needed to identify papers on the altmetrics platforms. For a more detailed description of data collection see Bar-Ilan et al. (2012).
[2] http://impactstory.org

Table 1. Coverage and citation or usage rates of a sample of 1,136 bibliometrics documents. "Events" are either bookmarks or citations, depending on the database.

|  | *Scopus* | *Web of Science* | *Mendeley* | *CiteULike* |
|---|---|---|---|---|
| Number of indexed documents | 1,136 | 957 | 928 | 319 |
| Total event counts | 18,755 | 17,858 | 8,847 | 777 |
| Percent sampled with nonzero event counts (total) | 85% (961) | 74% (845) | 82% (928) | 28% (319) |
| Mean events per article with nonzero count | 19.5 | 21.1 | 13.4 | 2.4 |

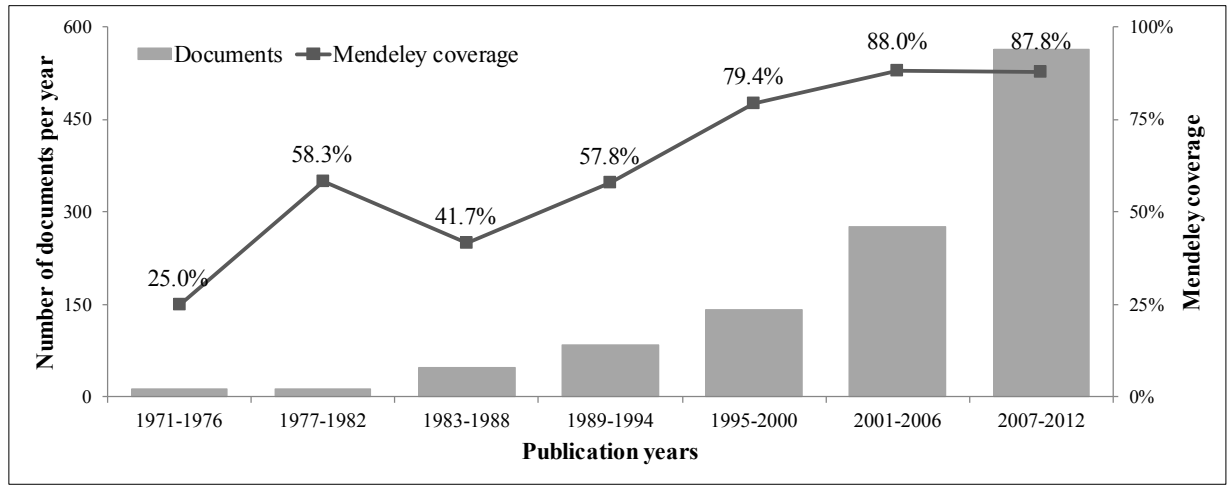

**Figure 1. Coverage of sampled documents in Mendeley per publication year. Overall coverage is 82% (*n=1,136*).**

### RQ 2: Use of Altmetrics Platforms by the Bibliometric Community

Since the results of RQ 1 confirmed that reference managers (Mendeley in particular) were a rich source for usage data and impact measurements of bibliometrics publications, we wanted to study who generates this usage data. To do this, we surveyed a sample of the bibliometrics community to learn how, when, and why they use various online environments; our goal was to better understand the significance of altmetrics indicators drawn from these environments.

*Method*

The paper and pencil survey was conducted among participants of the 17th International Conference on Science and Technology Indicators (STI) in Montréal. Participants filled out the survey during the conference from September 5th to 8th 2012. The survey contained open and closed questions; these mainly asked if and how members of the bibliometric community used social media with regards to organizing their literature and promoting their work, as well as how such tools influenced their professional lives. SPSS and Open Code were used for the analysis of the survey. All openly designed questions were coded using the Grounded Theory approach (Glaser & Strauss, 1967): codes were assigned to participants' statements, and these were then used to generate broader categories reflecting patterns of answering behavior.

*Results*

Of the 166 participants of the STI 2012 as indicated on the attendee list, 71 returned the questionnaire, resulting in a response rate of about 42.8%. Of the survey participants 63.4% were male and 33.8% were female, while 2.8% did not indicate their gender. Compared to the conference, females were somewhat overrepresented in our sample. While the youngest participant was 26 and the oldest 64, most respondents were between 31 and 40 years old. The mean age was 41.5 years. The respondents came from a mixed professional background, as 14.1% were research scientists and 14.1% worked in the R&D industry. 15.5% indicated that they had another background, 12.7% were doctoral candidates, 11.3% research managers, 8.5% government employees and 7.0% librarians. 4.2 % were associate professors/readers, 2.8% students, 2.8% postdocs, 2.8% assistant professors/lecturers and 2.8% full professors. One participant (1.4%) did not indicate his professional background.

Sixty people answered the question about reference management, 35 (58.3%) of whom use reference management software to organize scientific literature. The category "reference management software" includes desktop based software and web reference management services. A "personal solution" of literature management was described by 38.3% of respondents, which summarizes storing documents on personal drives on the desktop or on the Web as well as organizing literature on book shelves or in Word documents. Alerts from journals, bibliographic databases, or libraries fall in the category "information suppliers", which was described by 12 people (20.0%) as their way to find literature. Four people stated explicitly that they do not manage literature, because there is no need since they are not researchers.

When asked in a multiple choice question about whether they had heard of and used any of the social bookmarking services BibSonomy, CuL, Connotea, Delicious, or Mendeley, the latter was the most popular among respondents. Table 2 shows the percentage of the 70 respondents who knew and used the different bookmarking services and reference managers. Note that 77.1% of the respondents had heard about Mendeley, but only 25.7% actually used it. A similar percentage of the respondents had heard about CuL (72.9%), but only 12.9% of the respondents were actual users. The category "perceived usefulness" represents the percentage of a given platform's actual users compared to the number who have heard about it. By this measure, BibSonomy and CuL, were perceived to be relatively less useful; only 4.0% and 8.0% of those who knew the tools, respectively, actually use them. Mendeley was not only the most known tool, but also the one with the highest number of users. A third of all who had heard of the tool, used it, even though usage was rather occasional.

Table 2. Knowledge and usage of social bookmarking services and reference managers.

|  | BibSonomy | Connotea | CiteULike | Delicious | Mendeley |
|---|---|---|---|---|---|
| heard about the service *(n=70)* | 35.7% | 35.7% | 72.9% | 64.3% | 77.1% |
| used the service *(n=70)* | 1.4% | 2.9% | 12.9% | 11.4% | 25.7% |
| perceived usefulness | 4.0% *(n=25)* | 8.0% *(n=25)* | 17.6% *(n=51)* | 17.8% *(n=45)* | 33.3% *(n=54)* |

While there were more male than female users, the age structure of the Mendeley users corresponds to that of all participants. Both the youngest and the oldest respondent were Mendeley users. Although the numbers are too low to be representative, there seems a tendency towards a professional background in research of Mendeley users: the share of full professors, postdocs, doctoral candidates, and research scientists is higher among Mendeley users compared to the overall percentage of participants, while the percentage of research managers and members of R&D industry is lower. Thirteen of the 18 people who used Mendeley indicated for which purposes they used the tool. Managing references and connecting with people were equally important reasons to use Mendeley. This emphasizes that Mendeley connects literature management with the social aspect of connecting people who are interested in the same contents whereas CuL is mostly used for literature search.

The survey showed that Facebook, LinkedIn, Twitter, and Google+ were the most popular social networks. Figure 2 summarizes how many survey participants used the different social media tools. 52 people (73.2%) had a profile on Facebook, 48 (67.6%) on LinkedIn, 31 (43.7%) on Twitter, and 28 (39.4%) on Google+. Xing was used from 9.9% of users and 7.0% used MySpace. Among the tools focusing on the research community, Mendeley (23.9%), Academia.edu (21.1%), and ResearchGate (21.1%) have almost the same number of users in our sample, i.e. about one fifth of the participants had a profile on each of these platforms.

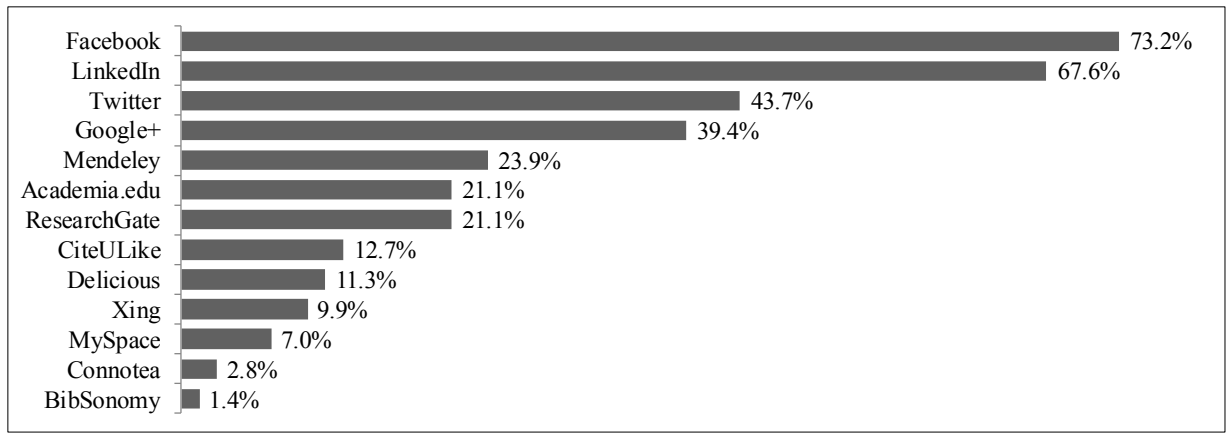

**Figure 2. Percentage of participants having a profile on or using social media tools mentioned in the survey (*n=71*).**

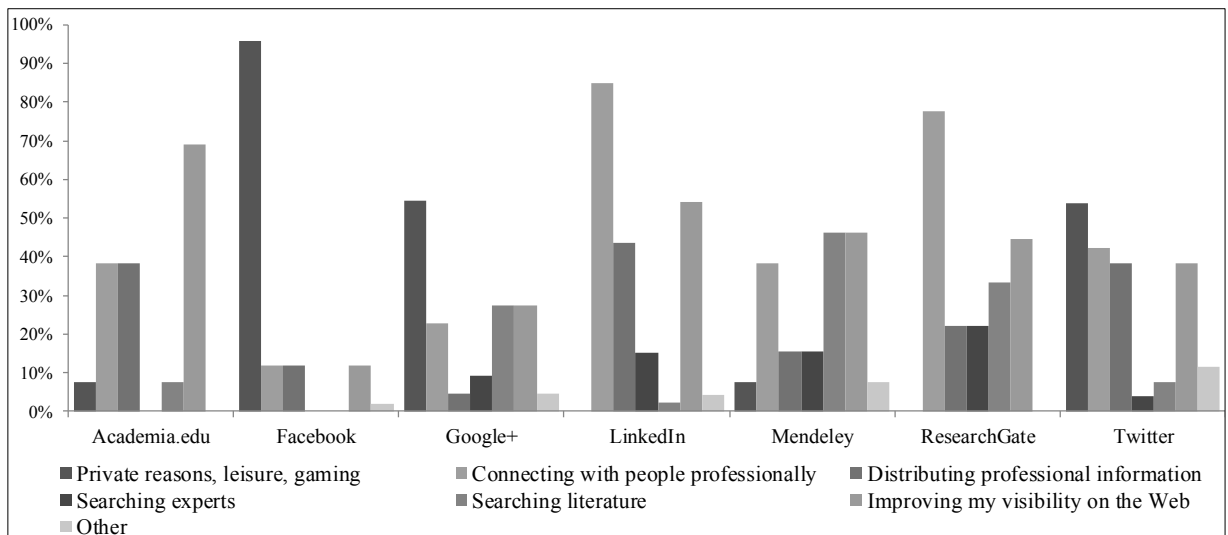

**Figure 3. What are participants using particular social networks for? Question allowed for multiple answers (Academia.edu: *n=13*; Facebook: *n=50*; Google+: *n=22*; LinkedIn: *n=46*; Mendeley: *n=13*; ResearchGate: *n=9*; Twitter: *n=26*. MySpace (n=4) and Xing (n=5) are not shown).**

Asking participants for the purpose of using these nine social networking platforms shows that Facebook, Google+, and MySpace are above all used for private purposes, while LinkedIn, ResearchGate, and Xing fulfill the main purpose of connecting with the professional community. LinkedIn is by far the most popular tool to connect with professional contacts; 84.8% indicated that this was the reason to use that platform. They also used LinkedIn to improve their own visibility (54.3%) and distribute professional information (43.5%). Twitter and Facebook were mostly used for private reasons, but Twitter was also important to connect with people professionally, distributing professional information and improving one's visibility. Although the overall use of Academia.edu was rather low (21.1% had a profile, but only 18.3% used it), 69.2% of the 13 Academia.edu users applied it to improve their visibility. Figure 3 shows the reasons for which respondents use social networks for each of the platforms.

Asked for personal publication profiles on Academia.edu, Google Scholar Citations, Mendeley, Microsoft Academic Search, ResearcherID (WoS), or ResearchGate, 32 participants listed their publications at least at one of these platforms. The most popular tool was Google Scholar Citations (22 respondents with profile; 68.8% of those with publication profiles), followed by ResearcherID (14: 43.8%), which can probably be attributed of the popularity and significance of Google and WoS. Google Scholar Citations (see Figure 4) was mostly used to check citations, WoS was used to check citations and add publications to the ResearcherID, while Academia.edu, Mendeley, and ResearchGate profiles were mostly used to add missing publications. In Microsoft Academic Search, people delete "wrong" publications from their profiles.

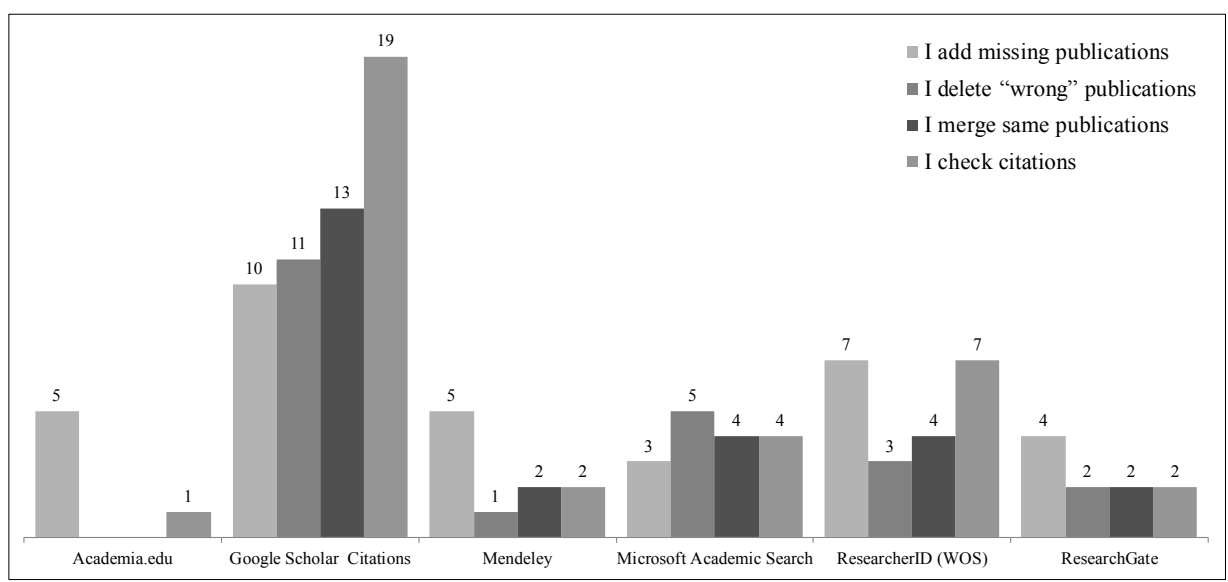

Figure 4. What are participants doing with their publication profile? Question allowed for multiple answers (Academia.edu: *n=5*; Google Scholar Citations: *n=22*; Mendeley: *n=8*; Microsoft Academic Search: *n=7*; Researcher ID (WoS): *n=14*; ResearchGate: *n=9*).

49.3% of the participants used some kind of repository to deposit their work. To 7 respondents the question did not apply, as they do not or no longer actively publish. Among those who used a repository, the most common was the institutional repository (57.1%), the second most popular was arXiv (21.4%). 47.9% of the respondents provided access to fulltexts on their homepages.

Although use of altmetrics platforms was quite low among survey participants, 85.9% thought that altmetrics had some potential in author or article evaluation. The majority, (71.8%) believed that the number of article downloads or views could be of use in author or article evaluation (see Figure 5 and Kurtz & Bollen, 2010 for a review of usage bibliometrics). Other sources such as citations in blogs (38.0%), Wikipedia links or mentions (33.8%), bookmarks on reference managers (33.8%), and discussions on Web 2.0 platforms (31.0%) were believed to have potential as altmetrics indicators as well.

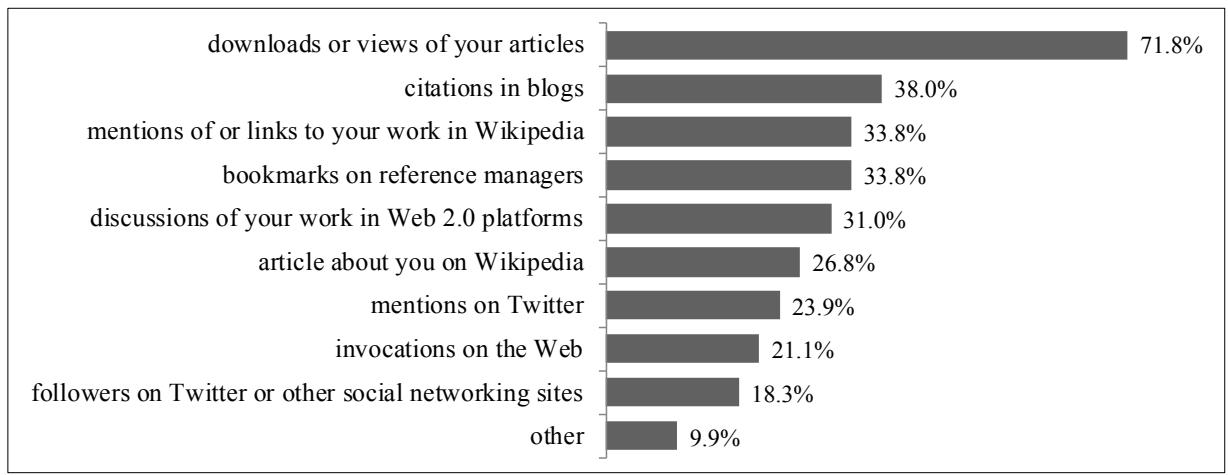

**Figure 5. Which alternative metrics are believed to have potential for article or author evaluation? Question allowed for multiple answers (*n=71*).**

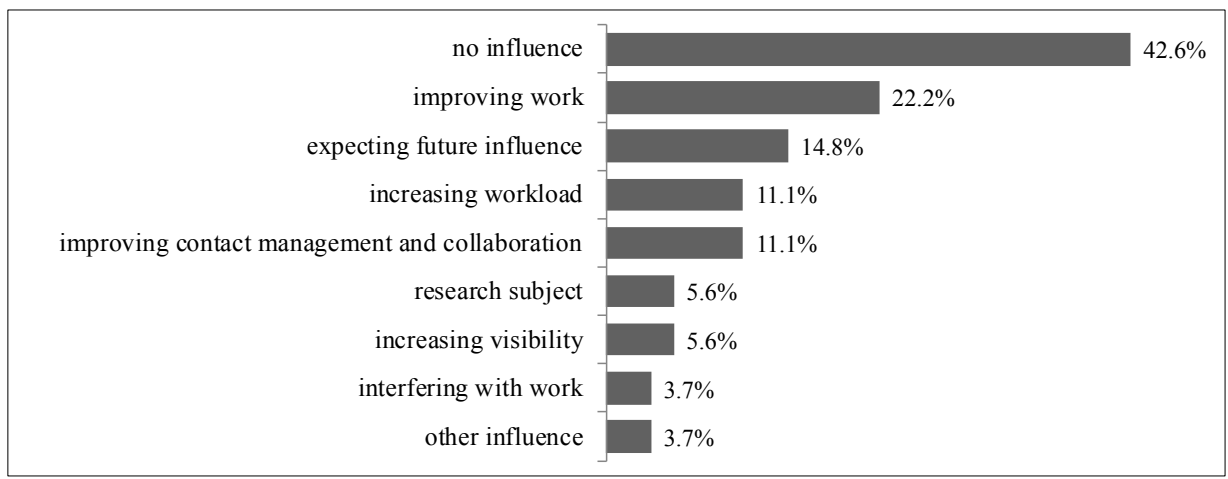

**Figure 6. In what ways do social network and bookmarking systems affect your professional life and/or work flow? Openly designed question (*n=54*).**

An openly designed question asked about in what ways social network and bookmarking systems affected their professional life and work flow (see Figure 6). Twenty-three (42.6%) of the 54 respondents said they were not at all influenced by these tools and 8 (14.8%) were not yet influenced but expected some impact in the future. 22.2% of respondents answered that the tools improved their work in terms of finding new information, fast distribution of information, and organization of research material. Two of these stated that social networks and social bookmarking systems "made my life much easier". For 11.1% the tools improved contact management and collaboration and 5.6% felt like they improved their visibility. On the other hand, 11.1% stated that social media tools increased their workload and 3.7% said that it interfered with their daily work, i.e. causing procrastination and getting lost in discussions on social media sites while delaying work.

**Conclusions and Outlook**

This study has followed a two-sided approach to explore the representativeness and validity of social media platforms to be used as data sources for altmetrics indicators evaluating impact of scholarly documents. It has shown that bibliometrics literature is well represented on social media platforms (i.e., Mendeley), making them a valuable source for evaluating the influence of scholarly documents in a broader way than citation analysis. The coverage of the sampled documents was as high as 82% overall with an even higher coverage of recent documents. Although this age bias was expected, as Mendeley was only launched in 2009, this bias needs to be considered when evaluating older documents. Mendeley did not only dominate in terms of coverage, but had also a much greater number of readers per document than CuL.

Having analyzed how bibliometrics documents are used on social reference managers, the second part of the study aimed to find out who was generating this use. A survey distributed among the core of the bibliometric community present at the 2012 STI conference in Montréal asked for social media use and its influence on the working environment of participants. Over half of those surveyed asserted that social media tools were affecting their professional lives, or that they were expecting future influence. Actual uptake of the platforms varied. Two-thirds of survey participants had LinkedIn accounts, which they used to connect professionally, while social networks with a scholarly focus such as Academia.edu, Mendeley, and ResearchGate were each used by only a fifth of respondents. Nearly half of those responding had Twitter accounts, which is extremely high compared to findings by Priem, Costello, and Dzuba (2011) and Ponte and Simon (2011), who found a Twitter usage rate of 2.5% and 18% among scholars, respectively; this may be due to growth in Twitter use, disproportionate use by bibliometricians, or the different methodologies employed.

Although Mendeley was the most popular social reference manager among the 71 participants, only one third surveyed use the tool, and and their use was rather sporadic. This is surprising given the high coverage of bibliometrics articles in Mendeley; it is unclear who is generating the high reader counts observed. A survey targeted directly at Mendeley users could clarify whether groups not at the conference (for example, people from other disciplines, or students, or practitioners) are using Mendeley heavily. The surveyed conference participants may also not properly represent the typical social media users and therefore reflect a biased picture of actual usage, although this assumption has to be proven in detailed studies. When altmetrics is broadly defined to include download data, 85% of bibliometricians surveyed expect at least one altmetrics indicator to become influential in future research evaluation. Around a third of respondents expected such influence from altmetrics based on blogs, Wikipedia, reference managers, and social media. Thus, although their use of social media tools remains modest as yet, survey participants are increasingly aware of the potential of altmetric indicators to supplement traditional evaluation indicators.

This study is limited by the specificity of its sample, and by potential non-response bias (enthusiastic users of social media may have been more likely to complete the survey). Results are thus not generalizable. Hence, further research should include the systematic analysis of all scholarly disciplines using this two-sided approach. Thus it would be possible to define the extent to which social media platforms cover a discipline's publication output as well as determine who is generating the use and for what purpose. This will help to validate altmetrics indicators as supplements to traditional metrics in research evaluation.